\documentclass[onecolumn]{revtex4}
\usepackage{graphicx}
\usepackage{color}

\newcommand{\rmd}{{\rm d}}
\newcommand{\rme}{{\rm e}}
\newcommand{\rmi}{{\rm i}}

\newcommand{\dt}[0]{\frac{{\rm d}}{{\rm d}t}}

\begin{document}

\title{High-performance solution of hierarchical equations of motions for studying energy-transfer in light-harvesting complexes}
\author{Christoph Kreisbeck}
\affiliation{Institut f\"ur Theoretische Physik, Universit\"at Regensburg, 93040 Regensburg, Germany}
\author{Tobias Kramer}
\email{tobias.kramer@mytum.de}
\affiliation{Institut f\"ur Theoretische Physik, Universit\"at Regensburg, 93040 Regensburg, Germany}
\author{Mirta Rodr{\'i}guez}
\affiliation{Instituto de Estructura de la Materia CSIC, C/ Serrano 121, 28006 Madrid, Spain}
\author{Birgit Hein}
\affiliation{Institut f\"ur Theoretische Physik, Universit\"at Regensburg, 93040 Regensburg, Germany}


\begin{abstract}
Excitonic models of light-harvesting complexes, where the vibrational degrees of freedom are treated as a bath, are commonly used to describe the motion of the electronic excitation through a molecule. Recent experiments point toward the possibility of memory effects in this process and require to consider time non-local propagation techniques. The hierarchical equations of motion (HEOM) were proposed by Ishizaki and Fleming to describe the site-dependent reorganization dynamics of protein environments (J.~Chem.~Phys., \textbf{130}, p.~234111, 2009), which plays a significant role in photosynthetic electronic energy transfer. HEOM are often used as a reference for other approximate methods, but have been implemented only for small systems due to their adverse computational scaling with the system size. Here, we show that HEOM are also solvable for larger systems, since the underlying algorithm is ideally suited for the usage of graphics processing units (GPU). The tremendous reduction in computational time due to the GPU allows us to perform a systematic study of the energy-transfer efficiency in the Fenna-Matthews-Olson (FMO) light-harvesting complex at physiological temperature under full consideration of memory-effects. We find that approximative methods differ qualitatively and quantitatively from the HEOM results and discuss the importance of finite temperature to achieve high energy-transfer efficiencies.
\end{abstract}

\maketitle

\section{Introduction}
Light-harvesting complexes (LHC) are pigment protein-complexes that act as the functional units of photosynthetic systems, capable of absorbing the energy of a photon and transferring it towards the reaction center where it is converted into chemical energy usable for the cell. The transfer of energy in such systems is described by electronic exciton-dynamics
coupled to the vibrations and other mechanical modes of the complex \cite{May2004a}. Laser spectroscopy shows quantum coherent effects in the energy transfer in LHC at temperatures up to $300$~K \cite{Engel2007a,Collini2010a,Panitchayangkoon2010a}.

Theoretical studies of model Hamiltonians at different levels of approximation \cite{Gaab2004a,Plenio2008a,Rebentrost2009a,Fassioli2010a,Wu2010a,Hsin2010,Hoyer2010a} show that the interplay between coherent transport and dissipation leads to high efficiencies in the energy transport in these systems. LHC provide a remarkable example of systems where noise or dissipation aids the transport. Understanding these systems is relevant as it gives insight into the optimal design of artificial systems such as novel nanofabricated structures for quantum transport or optimized solar cells.

The modelling of LHC is challenging due to the lack of atomistic ab-initio methods and requires to resort to effective descriptions. This is most apparent in the treatment of the vibrational excitations, which are commonly described by a structureless mode distribution. Then the energy transfer is calculated by the time propagation of a density matrix, which couples the electronic exciton dynamics to the vibrational environment. For LHC, the rearrangement of the molecular states after the absorption of the photon has to be taken into account and is described by the reorganization energy. The hierarchical equations of motion (HEOM) \cite{Yan2004a,Xu2005a,Ishizaki2005a} for the time evolution of the density-matrix were adapted by Ishizaki and Fleming \cite{Ishizaki2009c} to include the reorganization process in the transport equations and is exact within the model of exciton dynamics coupled to a bath with a Drude-Lorentz spectral density.

In principle the HEOM can be extended to other spectral densities by using a superposition of Drude-Lorentz peaks \cite{Meier1999a,Kleinekathofer2004}. Previous calculations for the energy-transfer efficiency of the FMO complex did not consider memory effects and used a weak coupling perturbation theory \cite{Rebentrost2009a,Fassioli2010a}. Other models try to get around these limitations by using the generalized Bloch-Redfield equations \cite{Wu2010a}, but  yield different results compared to the HEOM solution of the same model-system. Prolonged coherent dynamics is predicted due to the slow dissipation of reorganization energy to the vibrational environment \cite{Ishizaki2009a}. Theoretical descriptions must go beyond the rotating-wave approximation, perturbation theory, and require a full incorporation of time non-local effects, and physiological temperature. The HEOM fulfill all these premises.

To date, only the exciton population-dynamics for the FMO model has been studied within the full hierarchical approach \cite{Ishizaki2009a,Zhu2011a} whereas the calculation of efficiency or 2D  absorption spectra have been considered out-of-range for present computational power, since they require stable algorithms to propagate enlarged system matrices over many more time-steps.
The adverse computational scaling of the HEOM stems from the need to propagate a complete hierarchy of coupled auxiliary equations, which need to be simultaneously accessed in memory and propagated in time.
The insufficient computational power and memory-transfer bandwidth of conventional CPU clusters \cite{Struempfer2009a} has limited the application of the HEOM to study energy-transfer efficiency in small dimer systems, where other methods are available for comparison around $T=0$~K \cite{Anders2007a,Thorwart2009a,Roden2009,Prior2010a}. 
The advent of high-performance graphics processing units (GPU) with several hundred stream-processors working in parallel and with a high-bandwidth memory has lead us to perform the full HEOM approach for the exciton model of LHC. The efficiency calculations for the FMO system in the strong coupling regime require to propagate 240000 auxiliary matrices up to 50 ps (corresponding to 20000 time steps). The full HEOM approach takes only hours of computational time on a single GPU, whereas the corresponding CPU calculation would run several weeks and becomes completely unfeasible for bigger LHC due to the large communication overhead.
We use the GPU algorithmic advance to characterize the exciton energy-transfer efficiency in LHC for a wide range of reorganization energies under full consideration of the memory-effects and at $T=300$~K. Our calculations reveal several important mechanisms which are not contained within the approximative methods. The GPU-HEOM method opens the window to a wide-spread utilization of the HEOM, including the calculation of two-dimensional non-linear spectra of LHC as we will discuss elsewhere. Also the implementation of a scaled version of the HEOM \cite{Zhu2011a}, which reduces the number of auxiliary matrices, could be achieved on a GPU and reduces the computational effort of hierarchical methods further.

For the development of new theoretical chemistry and physics algorithms, GPU are important devices and considerably enlarge the class of solvable problems if one manages to devise a program code which takes full advantage of the GPU stream-processing architecture. For interacting many-body systems, this cannot be generally achieved by porting an existing program to the GPU, but requires to follow the vector-programming paradigm from the onset \cite{Olivares-Amaya2010a,Kramer2009c}.

The manuscript is organized as follows: in Sect.~\ref{sec:model} we set up the model for energy transfer to the reaction center in the FMO complex. In Sect.~\ref{sec:EffReorg} we calculate the key-quantities used to characterize  the 
energy flow, namely the efficiency and the transfer time to the reaction center. We compute them for a wide range of reorganization energies and bath correlation-times within the hierarchical approach. This section contains a detailed discussion of the differences of the HEOM results compared to calculations based on approximative methods. We highlight the main mechanism behind the high efficiency, the delicate balance between the requirements of an energy gradient towards the reaction center and the detuning of the energies, as shown in Sect.~\ref{sec:EffLevels}. In Sect.~\ref{sec:EffTemp} we discuss how the transport efficiency is optimized with respect to physiological temperature and comment on the thermalization properties of the HEOM. Finally we summarize our findings in Sect.~\ref{sec:conclusions}. Throughout the article, we provide detailed information about the computational times and requirements and collect in the appendices additional detailed information about the algorithms used and our GPU implementation.

\section{Model}\label{sec:model}

The FMO protein is part of the light harvesting complex that appears in green sulfur bacteria. Its structure has been widely studied both with X-ray and optical spectroscopic techniques \cite{Olson2004,Brixner2005,Milder2010a}. It has a trimer structure, with each of the monomers consisting of seven bacteriochlorophyll (BChl) pigment molecules, which are electronically excited when the energy flows from the antenna to the reaction center. An ab-initio calculation of the energy-transfer process within an atomistic model is far beyond present computational capabilities. Instead one has to develop effective model Hamiltonians such as the widely used excitonic Frenkel-Hamiltonian \cite{Leegwater1996a,Ritz2001a,May2004a}. Within the Frenkel model, which assumes that excitations enter the system one at a time, the seven BChl pigments of the FMO complex are treated as individual sites which are coupled to each other and also to the protein environment. The excitonic Hamiltonian is given by
\begin{eqnarray}
 \mathcal{H}_{\rm ex}&=& E_0 |0\rangle\langle0|
+\sum_{m=1}^N(\varepsilon_m^0+\lambda_m) |m\rangle\langle m|\nonumber \\ 
&&+\sum_{m>n}J_{mn} \left( |m\rangle\langle n|+|n\rangle\langle m| \right)
,\end{eqnarray}
where $N=7$, $|m\rangle$ corresponds to an electronic excitation of the chromophore BChl$_m$ and $|0\rangle$ denotes the electronic ground state of the pigment protein complex where we fix the zero of energy $E_0=0$. The site energies $\varepsilon_m=\varepsilon_m^0+\lambda_m$ of the chromophores consist of the ``zero-phonon energies'' $\varepsilon_m^0$ and a reorganization energy $\lambda_m$, which takes into account the rearrangement of the complex during excitation due to the phonon bath\cite{May2004a}
\begin{equation}
 \mathcal{H}_{\rm reorg}=\sum_{m=1}^N  \lambda_m |m\rangle \langle m|.
\end{equation}
In the following we will consider identical couplings for all sites, $\lambda_m=\lambda$.

\begin{figure}
\begin{center}
\includegraphics[width=0.5\columnwidth]{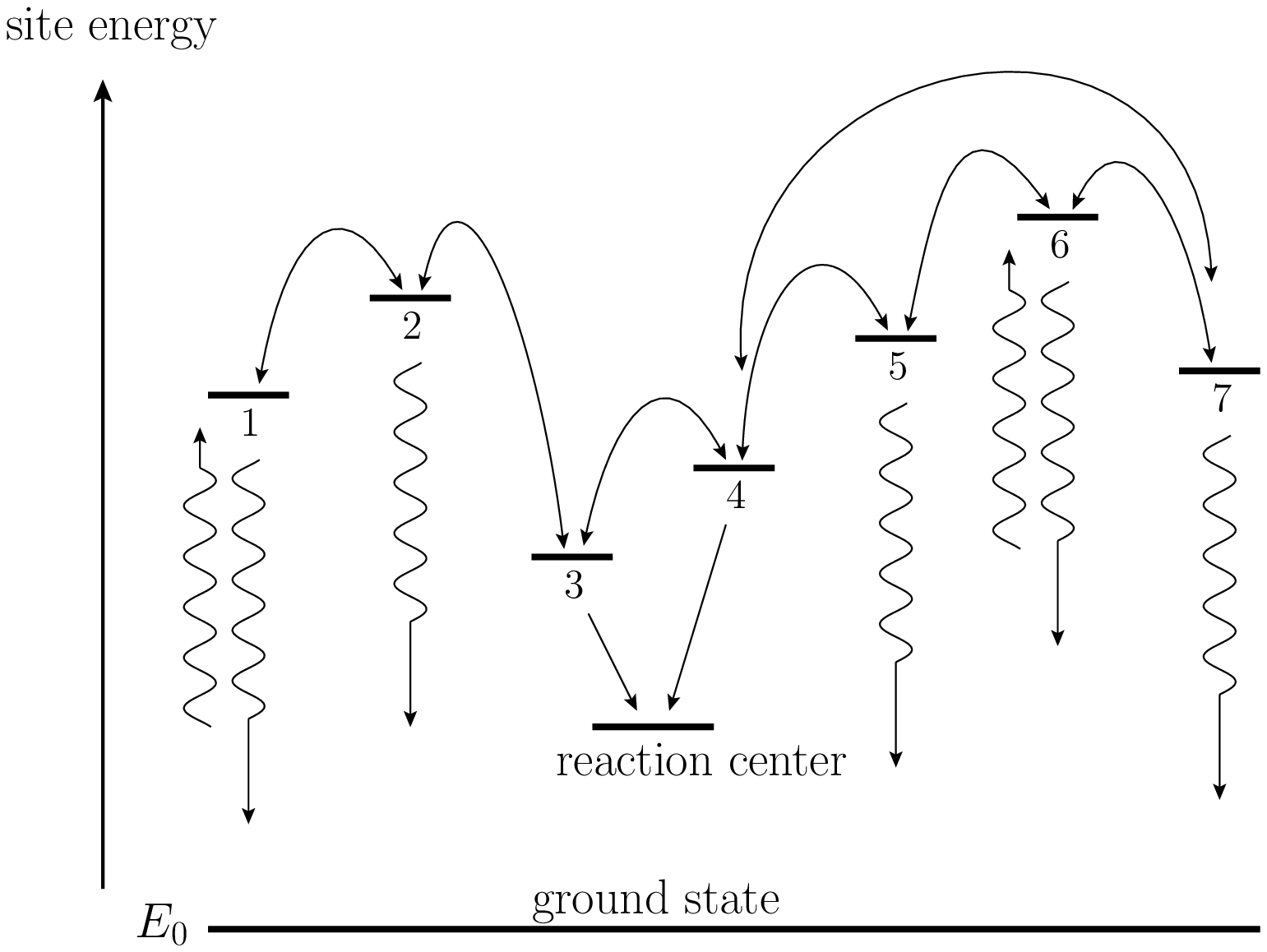}
\caption{\label{fig:sites} Sketch of the exciton energies of the FMO complex (\ref{tab:tab1}), the reaction center, and the ground state. Each site, designated with a number, represents a BChl pigment of the FMO complex. The arrows indicate the dominant inter-site couplings. The excitation enters the FMO complex through the chlorosome antenna located close to sites~1 and 6. The incoming excitation, depicted with wavy arrows pointing upwards, follows two energy pathways to the reaction center. Wavy arrows pointing downwards indicate radiative loss-channels leading to the electronic ground state. In addition, each site is coupled to a phonon bath which accounts for the protein environment surrounding the pigments.}
\end{center}
\end{figure}

The inter site couplings $J_{mn}$ are obtained by fits to experimentally measured absorption spectra~\cite{Milder2010a}. In this contribution we use the designations and parameters of Ref.~\cite{Renger2006a}, table~4 (trimer column) and table~1 (column 4), summarized in \ref{tab:tab1}.
A sketch of the dominant couplings is shown in \ref{fig:sites}.
\begin{table}[b]
\begin{center}
\begin{tabular}{c c c c c c c c}
\hline
                   & BChl$_1$ & BChl$_2$ & BChl$_3$ & BChl$_4$ & BChl$_5$ & BChl$_6$ & BChl$_7$ \\
\hline
BChl$_1$ &    \textbf{12410}  &  \textbf{-87.7}   &   5.5                    &      -5.9                 &            6.7            &         -13.7           &         -9.9               \\
BChl$_2$ &     \textbf{-87.7}   & \textbf{12530}  & \textbf{30.8}    &    8.2                     &         0.7               &        11.8             &       4.3                  \\
BChl$_3$ &                    5.5       &  \textbf{30.8}    &  \textbf{12210} & \textbf{-53.5}    &        -2.2               &            -9.6          &        6.0                  \\
BChl$_4$ &            -5.9              &        8.2                &  \textbf{-53.5}   & \textbf{12320}  &   \textbf{-70.7}  & -17.0                  &    \textbf{-63.3}   \\
BChl$_5$ &            6.7               &       0.7                 &       -2.2                 &  \textbf{-70.7}  & \textbf{12480} & \textbf{81.1}     &          -1.3                \\
BChl$_6$ &          -13.7              &           11.8          &        -9.6                &           -17.0         &  \textbf{81.1}   & \textbf{12630}  &      \textbf{39.7}    \\
BChl$_7$ &            -9.9              &              4.3          &      6.0                   &   \textbf{-63.3}  &            -1.3          & \textbf{39.7}    &   \textbf{12440}   \\
\end{tabular}
\caption{\label{tab:tab1} Exciton Hamiltonian in the site basis in (cm$^{-1}$). Bold font shows the dominant couplings and site energies. Values taken from Ref.~\cite{Renger2006a}.}
\end{center}
\end{table}
The protein environment surrounding the pigments is modeled as identical featureless spectral bath densities coupled to each BChl. For simplicity, we neglect correlations between the baths. The electronic excitations at each site couple linearly with strength $d_i$ to the vibrational phonon modes 
 $b_i^\dag$ of frequency $\omega_i$. The coupling Hamiltonian is given by
\begin{eqnarray}
 \mathcal{H}_{\rm ex-phon}&=&\sum_{m=1}^N \left( \sum_i \hbar \omega_i d_i(b_i+b_i^\dag)\right)_m   |m\rangle\langle m|,
\end{eqnarray}
where we assume identical baths at every site. Note that the reorganization energy is related to the coupling by
$\lambda=\sum_i\hbar\,\omega_i\,d_i^2/2$.

We model the losses due to radiative decay from the exciton to the electronic ground state $|0\rangle$ introducing a dipole coupling to an effective radiation photon field $a_{\nu}^\dag$
\begin{equation}
\mathcal{H}_{\rm ex-phot}=
\sum_{m=1}^N\sum_{\rm \nu}  (a_{\nu}+a_{\nu}^\dag)\mu^\nu_m
\left( |0\rangle\langle m| +|m\rangle\langle 0|\right), \label{eq:exphot}
\end{equation}
which results in a finite life-time for the exciton. The reaction center (RC) is treated as a population-trapping state 
\begin{equation}
\mathcal{H}_{\rm trap}=E_{RC}|RC\rangle\langle RC|
\end{equation}
and enlarges the system Hamiltonian to a $9\times 9$ matrix. 
Adolphs and Renger \cite{Renger2006a} suggest that pigments~$3$ and $4$, which have
the largest overlap with the energetically lowest exciton-state, couple to the reaction
center. Recent experimental evidence shows that pigment~$3$ is orientated towards the reaction center \cite{Wen2009a}. In addition it has been proposed that
an 8th pigment may play a role in the initial stages of the energy transfer \cite{Schmidt2011a}.
Here, we include the reaction center by introducing leakage rates from pigments~$3$ and $4$ to the reaction center, which acts as a population trapping state.
Thus the coupling term to the reaction center reads
\begin{equation}
\mathcal{H}_{\rm ex-RC}=\sum_{m=3}^4\sum_{\rm \nu'}  (a_{\rm \nu'}+a_{\rm \nu'}^\dag)
\mu_{RC}^{\nu'} \left( |RC\rangle\langle m|+|m\rangle\langle RC| \right) \label{eq:exrc}
\end{equation}
where the sum runs over the photon modes at the reaction center. As shown in Sect.~\ref{sec:markovif}, Eqs.~(\ref{eq:lab2},\ref{eq:lab3}), the coupling can be expressed in terms of a trapping rate $\Gamma_{\rm RC}$, and similarly for the radiative decay in \ref{eq:exphot} with the rate $\Gamma_{\rm phot}$.
The total Hamiltonian of the system is thus given by
\begin{eqnarray}
 \mathcal{H}&=& \mathcal{H}_{\rm ex}+ \mathcal{H}_{\rm trap}  + \mathcal{H}_{\rm ex-phon}
+\mathcal{H}_{\rm ex-phot}\nonumber\\
&&+\mathcal{H}_{\rm ex-RC}
+ \mathcal{H}_{\rm phon}+ \mathcal{H}_{\rm phot}^0+ \mathcal{H}_{\rm phot}^{\rm RC},
\end{eqnarray}
where $\mathcal{H}_{\rm phon}= \sum_{i,m}  (\hbar \omega_i b^\dag_i b_i)_m $,
$\mathcal{H}_{\rm phot}^0=\sum_{\nu,m} (  h\nu  a^\dag_{\nu} a_{\nu})_m$, and
$\mathcal{H}_{\rm phot}^{\rm RC}= \sum_{\nu',m=3,4}  ( h\nu'  a^\dag_{\nu'} a_{\nu'})_m$.
The time evolution of the total density operator $R(t)$ is described by the Liouville equation
 \begin{equation}\label{eq:Liouvaa}
  \dt R(t)=-\frac{\rmi}{\hbar}[\mathcal{H},\ R(t)]. \label{eq:L}
  \end{equation} 
We assume that at initial time $t=0$ the total density operator factorizes in system and bath components
\begin{equation}
R(t=0)=\rho(t=0)\otimes\rho_{\rm phon}\otimes\rho_{\rm phot}^0\otimes\rho_{\rm phot}^{\rm RC} \label{eq:R},
\end{equation}
while at later times the system and the bath get entangled.
Since we are only interested in the exciton dynamics, we trace out the degrees of freedom of the phonon and photon environments $\alpha=\{\rm phon, phot^0, phot^{RC}\}$
and propagate the reduced $9 \times 9$ density matrix in the Schr\"odinger picture
\begin{eqnarray}\label{eq:10aa}
\rho(t)&=&
\mbox{Tr}_\alpha\big( 
{\rm e}^{-\frac{\rmi t}{\hbar}(
 \mathcal{L}_0
+{\mathcal{L}}_{\rm ex-phon}
+{\mathcal{L}}_{\rm ex-phot}
+{\mathcal{L}}_{\rm ex-RC} 
+{\mathcal{L}}_{\rm bath} 
)
} R(0)\big)
\end{eqnarray}
for the exciton system $\{|m\rangle\}_{m=1,\ldots,7}$, the ground electronic state $|0\rangle$, and the reaction center $|{\rm RC}\rangle$. 

Eq.~(\ref{eq:10aa}) is obtained by formal integration of the Liouville equation~(\ref{eq:Liouvaa}).
The operator $\mathcal{L}_{0}=[\mathcal{H}_{\rm ex}+ \mathcal{H}_{\rm trap} ,\bullet]$ represents the coherent dynamics and $\mathcal{L}_{\rm ex-phon}$ accounts for dephasing and energy relaxation due to vibrations induced by the interaction with the protein environment, while the recombination and energy trapping are expressed by $\mathcal{L}_{\rm ex-phot}$ and $\mathcal{L}_{\rm ex-RC}$, respectively. The parts describing the different baths are summarized in $\mathcal{L}_{\rm bath}=[ \mathcal{H}_{\rm phon}+ \mathcal{H}_{\rm phot}^0+ \mathcal{H}_{\rm phot}^{\rm RC},\bullet]$.
The coupling to the phonon and photon baths can be studied with different degrees of approximation.

We calculate the energy flow within a hybrid formulation which treats the exciton dynamics and the vibrational environment within the HEOM and the trapping to the reaction center and the radiative decay within a Markov model. The Markovian treatment of the photon modes is justified as it occurs in a very different time scale and no backward energy flow to the system is allowed. We abbreviate our model by ME-HEOM, see Sect.~\ref{sec:markovif}. We solve the hierarchical equations using GPUs, which are ideally suited for this task and lead to huge speed-ups of the algorithm. Details of the computational implementation are collected in Sect.~\ref{sec:gpu}.

\section{Trapping time for different reorganization energies}\label{sec:EffReorg}

The strong coupling of the excitonic system to the vibrational environment, which is of the same order as the excitonic energy differences (100~cm$^{-1}$), requires a detailed treatment of the phonon bath over the time-scale of the correlations present in the system. The coupling is quantified by the parameter $\gamma$ Eq.~(\ref{eq:gamma}), ranging from (35-166~fs)$^{-1}$ for models of light-harvesting complexes \cite{Ishizaki2009a}.
We calculate the efficiency of the energy transfer from an initially excited site to the reaction center using the hierarchical equations (\ref{eq:labsevena},\ref{eq:lab8}). The efficiency $\eta$ is defined as the population of the reaction center at long times
\begin{equation}\label{eq:eta}
\eta=\langle RC | \rho  (t\rightarrow\infty) | RC \rangle.
\end{equation}
For the FMO complex, two sites are located near the light-absorbing antenna  \cite{Renger2006a}. We consider initial excitations at either site~1 or 6, which give rise to two energy pathways to the reaction center. One pathway starts from site~1 and transfers energy via site~2 to site~3, and the second pathway starts from site~6 and the energy flows via site~7 or 5 to site~4, see \ref{fig:sites}.

We fix the upper limit of time propagation at $t_{\rm max}$, defined such that the remaining population in the system, excluding the ground-state and reaction center, has dropped from initially $1$ to $10^{-5}$. 
To our knowledge, no solid experimental data exists for the coupling strength in eq.~(\ref{eq:exrc}), given
in terms of the trapping rate $\Gamma_{\rm RC}$ of sites~3 or 4 to the reaction center. In
the following we assume values of $\Gamma_{\rm RC}^{-1}=2.5$~ps and $\Gamma_{\rm phot}^{-1}=250$~ps, which are of the same order of magnitude as in other theoretical studies \cite{Rebentrost2009a,Hoyer2010a,Wu2010a}.
\begin{figure}
\begin{center}
\includegraphics[width=0.5\columnwidth]{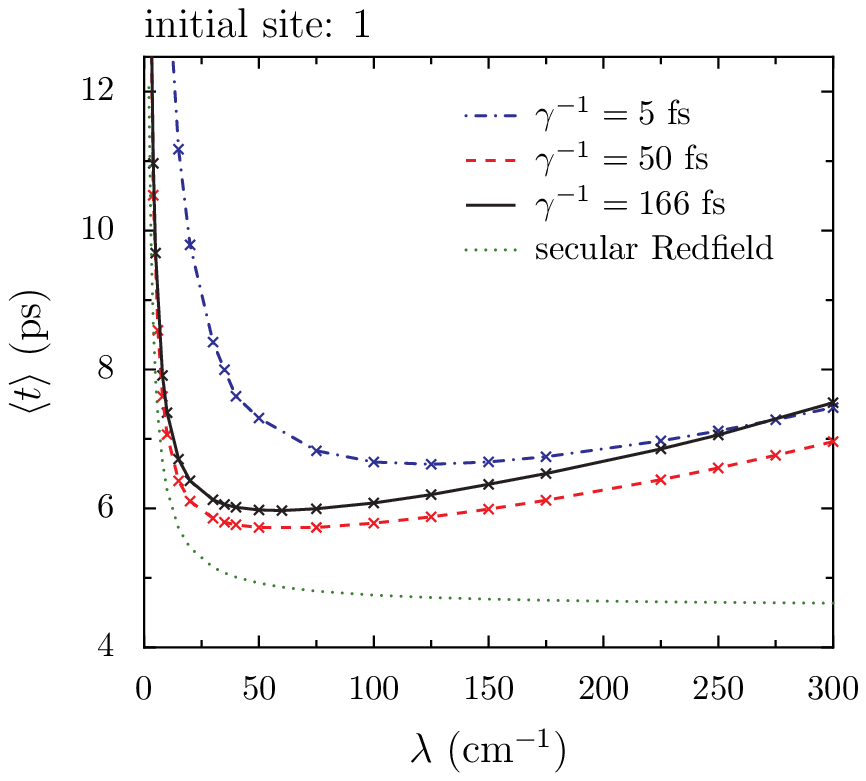}
\includegraphics[width=0.5\columnwidth]{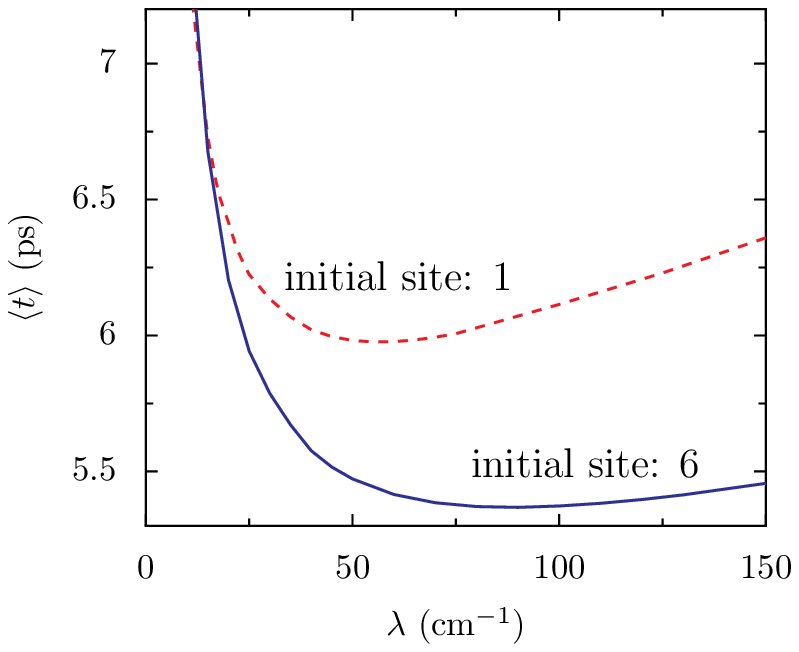}
\caption{\label{fig:reorg}
Trapping time from \ref{eq:trappingtime} as function of reorganization energy $\lambda$ at temperature $T=300$~K. Trapping rate to BChl $3$ and $4$ $\Gamma_{\rm RC}^{-1}=2.5$~ps and $\Gamma_{\rm phot}^{-1}=250$~ps.
Upper panel: secular Redfield result with $\gamma^{-1}=166$~fs and the ME-HEOM results for three different bath correlation times $\gamma^{-1}=166$~fs, $\gamma^{-1}=50$~fs, $\gamma^{-1}=5$~fs. 
The excitation enters at site 1.
Lower panel: Comparison of the trapping times for the two possible pathways in the FMO when the energy is entering the complex starting from site~1, or at site 6 for a bath correlation time of $\gamma^{-1}=166$~fs.
}
\end{center}
\end{figure}
The actual time scale of the energy trapping is quantified by the trapping time
\begin{equation}\label{eq:trappingtime}
\langle t \rangle=\int_0^{t_{\rm max}} \rmd t'\ t'\, \big(\frac{\rmd}{\rmd t} \langle RC | \rho(t) | RC \rangle \big)_{t=t'},
\end{equation}
where we replace the upper limit of the integral by $t_{\rm max}$.
The trapping time depends strongly on the reorganization energy as shown in \ref{fig:reorg}. For reorganization  energies $\lambda<50$~cm$^{-1}$ the coupling to the environment assists the transport and the trapping time decreases when $\lambda$ increases. 

Evaluating the equations of motion (\ref{eq:labsevena},\ref{eq:lab8}) in the ME-HEOM approach requires to truncate the hierarchy at $N_{\rm max}$, which has to be large enough to reach convergence. 
In \ref{fig:reorg} we adjust the truncation such that the trapping times for $N_{\rm max}=N$ and $N_{\rm max}=N+1$ differ at most by 0.02~ps. The required truncation increases with reorganization energy and for $\lambda=300$~cm$^{-1}$ we need $N_{\rm max}=16$ where we have to propagate 245157 auxiliary matrices over 22000 time steps ($\Delta t$=2.5~fs) leading to a GPU computation time of 3.7 hours. On a standard CPU the same calculation takes more than one month and a systematic study of parameters is not feasible.

In the upper panel of \ref{fig:reorg} we compare the ME-HEOM result with the secular Redfield theory, which employs the time-local Born-Markov approximation in combination with the rotating-wave approximation. For stronger values of the coupling, the hierarchical approach strongly deviates from the plateau obtained within the secular Redfield theory, which assumes a fast decay of the phonon bath. The secular Redfield limit (see Sect.~\ref{sec:markovif}) reflects, as expected, the qualitative behavior only for small reorganization energies and overestimates the energy transfer to the reaction center
for $\lambda>10$~cm$^{-1}$.

An interesting question is the existence of an optimal value for the coupling $\lambda$ and the bath correlation-rate $\gamma$, for which the trapping time is minimized (and the efficiency maximized). Secular and full Redfield do not yield a local minimum of the trapping time, and thus no corresponding optimal $\lambda$. Introducing the bath-correlations and memory effects by the parameter $\gamma$ in the ME-HEOM gives rise to a local minimum and an optimal value of $\lambda$, as shown in \ref{fig:reorg}. In addition an optimal value of $\gamma$ emerges around $\gamma^{-1}=25-35$~fs. For a small value $\gamma^{-1}=5$~fs, the theory predicts a rapid loss of efficiency. 

The lower panel of \ref{fig:reorg} details the changes of the trapping time for the two different pathways of the energy flow in the FMO complex as function of the reorganization energy. The optimal reorganization energy for an initial excitation of site~1 is given by $\lambda_{\rm opt}^{1}=55$~cm$^{-1}$ ($\langle t\rangle_{\rm opt}^{1}=6.0$~ps), while for an initial excitation of site~6  we obtain $\lambda_{\rm opt}^{6}=85$~cm$^{-1}$ ($\langle t\rangle_{\rm opt}^{6}=5.4$~ps). 

Optimal values of trapping times have been calculated within the generalized Bloch-Redfield (GBR) approximation \cite{Wu2010a}. Using the same parameters, couplings, and Hamiltonian as in Ref.~\cite{Wu2010a}, the ME-HEOM yield qualitative and quantitative differences with a $0.9$~ps longer trapping time for an initial excitation of site $1$. For an initial excitation located at site $6$ the ME-HEOM and GBR results for the trapping time differ by $0.2$~ps.

\section{Efficiency for rearranged energy levels}\label{sec:EffLevels}
\begin{figure}
\begin{center}
\includegraphics[width=0.5\columnwidth]{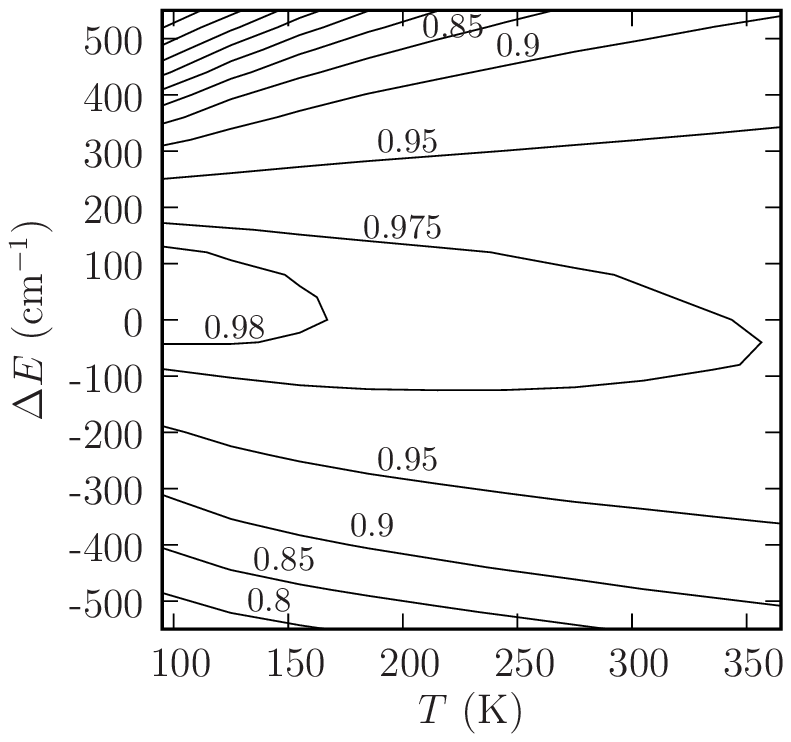}
\includegraphics[width=0.5\columnwidth]{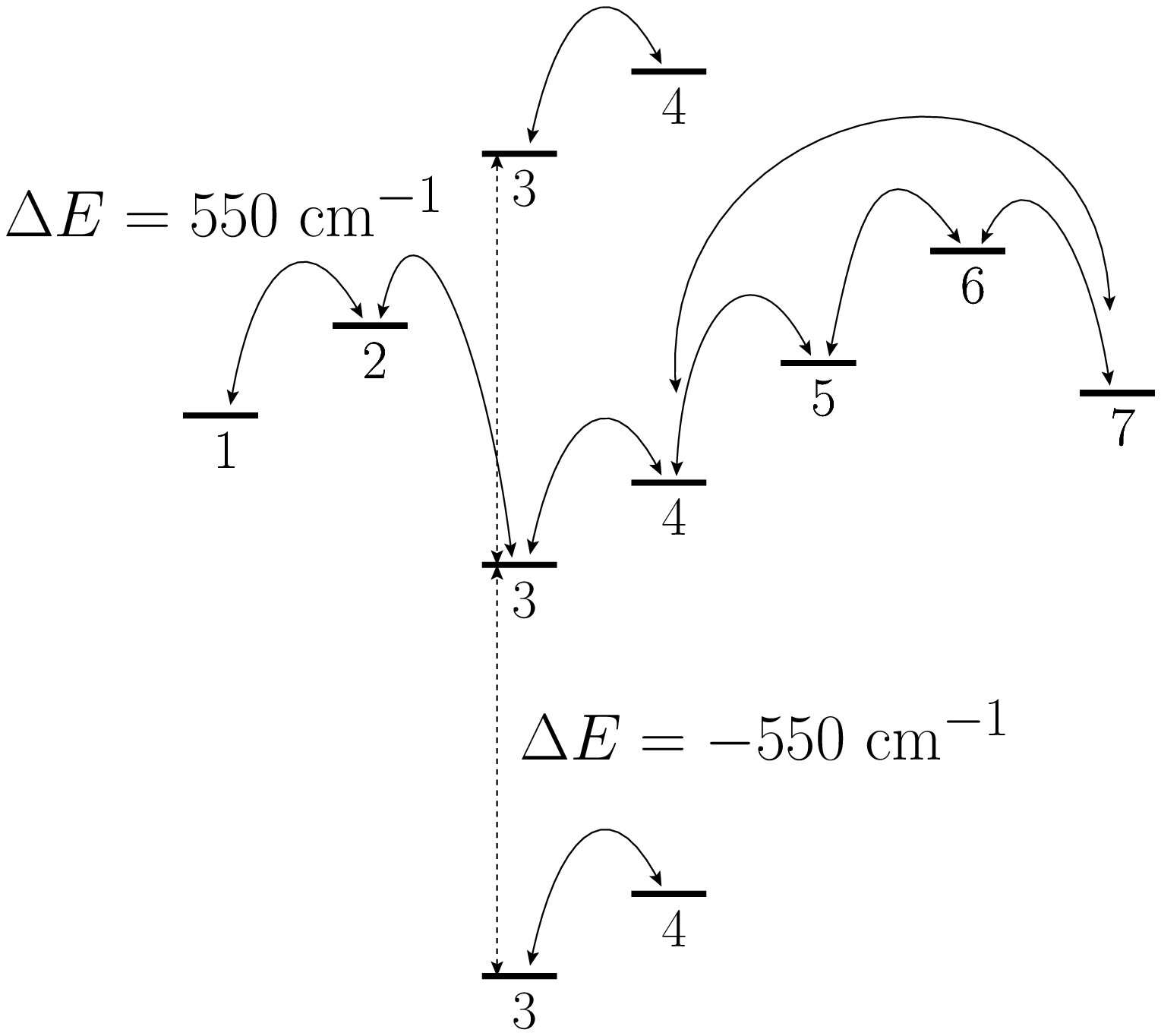}
\caption{\label{fig:efflevelshift}Upper panel: Energy transfer efficiency $\eta$ in \ref{eq:eta} as function of temperature and site-energy shifts $\varepsilon_{3/4}\rightarrow\varepsilon_{3/4}+\Delta E$. ME-HEOM parameters: $\lambda=35$~cm$^{-1}$, $\gamma^{-1}=166$~fs, $\Gamma_{\rm phot}^{-1}=250$~ps and $\Gamma_{\rm RC}^{-1}=2.5$~ps. The hierarchy is truncated at $N_{\rm max}=8$.  Lower panel: energy-level shifts considered in the parameter range of the left panel.
}
\end{center}
\end{figure}

In this section we study the relevance of the spacings of the energy levels in the FMO complex to see if the experimentally obtained energy levels (\ref{tab:tab1}) are close to an optimal value with respect to transport efficiency at physiological temperature. 

The isolated excitonic system shows coherent oscillations of energy between the initially populated site and the delocalized excitonic states. Coupling to the environment gives rise to several mechanisms leading to a non-reversible energy transfer.
In the simplest Haken-Strobl model, only dephasing is incorporated \cite{Rebentrost2009a,Chin2010a}, but the temperature is fixed at $T=\infty$. Only by adjusting the dephasing rate, temperature effects can be included on a rudimentary level. The ME-HEOM approach enables us to calculate the transport at physiological temperature ($T=300$~K) and brings into the picture another crucial mechanism to achieve highly efficient energy transfer. Namely, the temperature dependent stationary site populations. Since the system is in contact with a thermal environment at finite temperature, there is energy dissipation and the system relaxes to thermal equilibrium. This process guides the excitons to the lowest energy states (for the FMO complex within a few picoseconds) and is not contained in pure dephasing models.

For a small coupling $\lambda$ and under the assumption that the system and bath degrees of freedom factggze, the thermal state of the system is given by the Gibbs measure 
\begin{equation}
 \rho_{\rm thermal}=e^{-\beta \mathcal{H_{\rm ex}}}/\mbox{Tr}\,e^{-\beta \mathcal{H_{\rm ex}}},\ \beta=1/(k_B T),
\end{equation}
which populates the eigenstates of $\mathcal{H}_{\rm ex}$ according to the Boltzmann statistics. Stronger couplings lead to deviations from the Boltzmann statistics \cite{Zuercher1990a}.
Since the coupling to the reaction center, where the system deposits its excitation, is linked to sites $3$ and $4$, the efficiency 
depends strongly on the population and actual site-energies $3$ and $4$.
To study this relation, we shift levels  $\varepsilon_{3/4}\rightarrow\varepsilon_{3/4}+\Delta E$ and compute the efficiency of the energy transfer. \ref{fig:efflevelshift} shows the efficiency evaluated with the ME-HEOM. We observe an almost symmetric behavior of the efficiency for positive and negative energy shifts, with slightly higher efficiencies towards negative energy shifts. 

A shift to lower energies increases the energy gradient in the FMO as the thermal state prefers to populate the low-lying sites. This mechanism improves the transfer efficiency but shifts the two sites out of resonance and they get decoupled from the other levels of the FMO. Thus coherent transport becomes more difficult and the energy transfer to the reaction center is expected to slow down.
Similar arguments hold when the energies $\varepsilon_{3}$ and $\varepsilon_{4}$ are shifted to higher energies.
On the one hand $\Delta E>0$ brings the sites~3 and 4 closer to resonance and increases the coupling to the remaining sites, thus enhancing coherent transport. On the other hand the thermal state gets delocalized over all sites of the FMO complex  and there is no special preference to
populate site~$3$ and site~$4$. In such case  the FMO loses its property to act as an energy funnel and environment assisted transport
to the reaction center is hindered.

\ref{fig:efflevelshift} shows how the delicate interplay between coherent delocalization and energy dissipation towards the reaction center gives rise to an optimal arrangement of site energies. We obtain maximal efficiency around $\Delta E=0$ corresponding to the original parameters in \ref{tab:tab1} and the optimum value is robust against small variations in the site energies.

\section{Trapping time for different temperatures}\label{sec:EffTemp}
\begin{figure}
\begin{center}
\includegraphics[width=0.45\columnwidth]{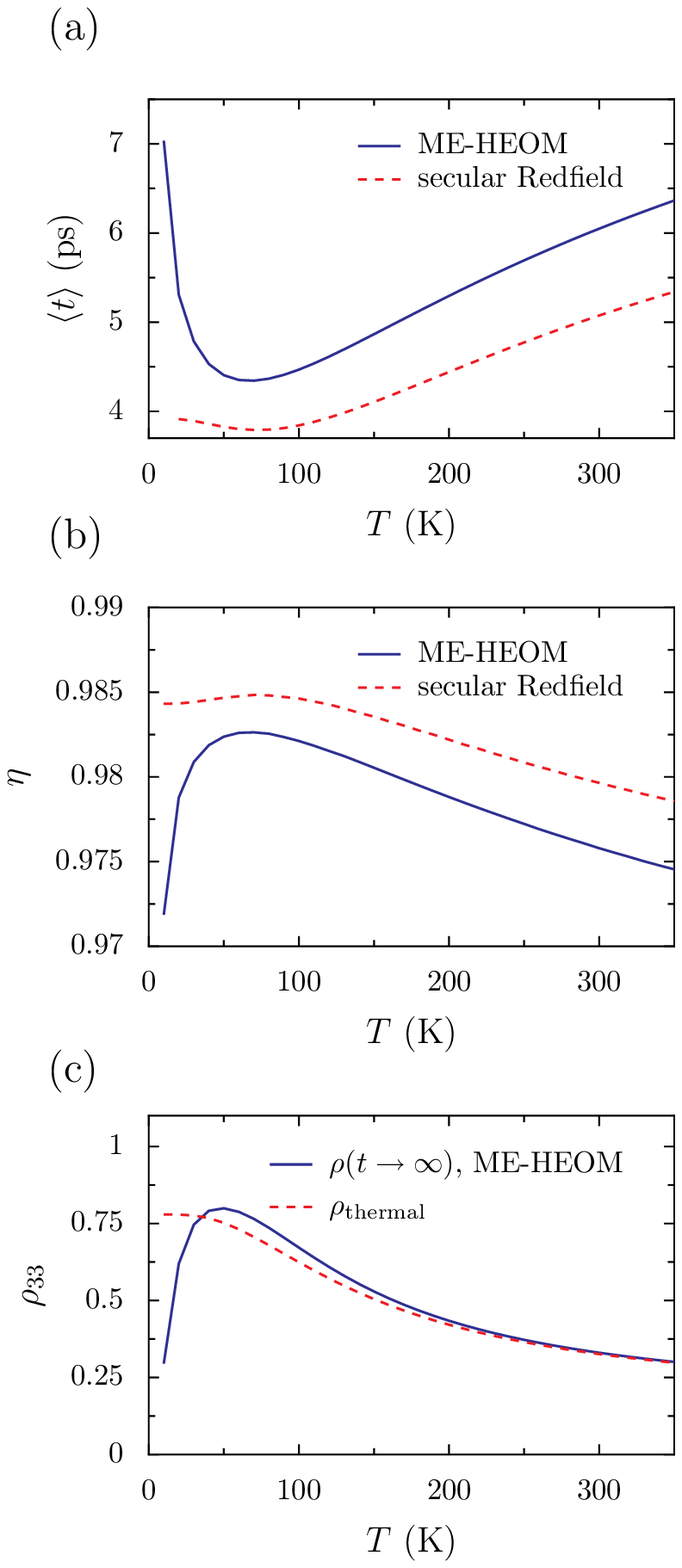}
\caption{\label{fig:efftemp}
Energy transfer as a function of temperature for the secular Redfield approximation and the exact ME-HEOM calculation with $\gamma^{-1}=166$~fs, $\Gamma_{\rm phot}^{-1}=250$~ps, $\Gamma_{\rm RC}^{-1}=2.5$~ps and truncation $N_{\rm max}=8$. Both approaches use a reorganization energy of $\lambda=$35~cm$^{-1}$ and start with initial population at site~$1$.
(a) Trapping time as a function of temperature.
(b) Efficiency as a function of temperature.
(c) Population of site $3$ for different temperatures in the Boltzmann thermal equilibrium state $\rho_{\rm thermal}$ and for the isolated FMO (decoupled from the reaction center and the radiative decay) using the ME-HEOM, $\rho(t\rightarrow\infty)$. Note that the ME-HEOM needs further corrections at temperatures below $100$~K in order to reach the thermal state.
}
\end{center}
\end{figure}
As discussed in the previous section, the environment assists the transport towards the thermal equilibrium state. In the FMO complex, the sites $3$ and $4$ are coupled to the reaction center and present the lowest exciton energies in the system (see 
\ref{fig:sites}), thus the energy dissipation in the phonon environment enhances the population of those sites and hence the efficiency.With increasing temperature one might expect high transfer efficiencies because thermalisation occurs on a faster time scale. However, with increasing temperature higher energy states have a higher thermal-equilibrium population and thus the transport efficiency towards the reaction center decreases. 

These two competing mechanisms result in an optimal temperature with maximal efficiency.
Both mechanism are already present in the secular Redfield limit, and the optimal energy transfer is obtained around $75$~K, see \ref{fig:efftemp}(a). 
Our ME-HEOM calculations predict optimal efficiency at slightly lower temperature $70$~K, but this value is outside the range where our high-temperature implementation is supposed to work (see Sect.~\ref{sec:markovif}). We obtain a steep increase of the trapping time for low temperatures shown in \ref{fig:efftemp}(a), which is also reflected in the efficiencies \ref{fig:efftemp}(b). 
This increase in trapping time and decrease in efficiency is not present in the secular Redfield approach, which saturates for $T\rightarrow 0$. Although we take into account the lowest-order quantum correction to the Boltzmann statistics \cite{Ishizaki2009a}, at low temperatures more correction terms are required. One criteria to validate the HEOM is to check if the stationary state $\rho(t\rightarrow\infty)$ of the population dynamics of the isolated FMO, which is decoupled from the reaction center and radiative decay, approaches the thermal state.
As is shown in \ref{fig:efftemp}(c), the HEOM high-temperature implementation fails to approach the thermal state
for temperatures below $100$~K, where the HEOM predict an unphysical steep decent of population at low energy site~3 and hence
transfer efficiency is underestimated. 
For temperatures above $100$~K the high temperature limit agrees very well with the thermal state and the ME-HEOM results are reliable.
Comparing our ME-HEOM results above $100$~K to the secular Redfield ones shown in \ref{fig:efftemp}(a) and (b) we conclude that the Redfield approach, which is known to be valid in the weak coupling limit only, overestimates the efficiency and underestimates the trapping time.

\section{Conclusions}\label{sec:conclusions}

We have shown that the HEOM are computationally feasible for calculating the energy transfer for large systems following our GPU implementation. This algorithmic advance allowed us to calculate the efficiency and trapping time of the energy transfer in the FMO complex for a wide range of parameters. 
The results point to qualitative and quantitative deficiencies of approximative methods and show that an accurate treatment of memory effects and reorganization processes in the system-bath coupling of LHC is needed to evaluate the precise role of temperature, exciton energy-differences, the coupling strength, and the time correlations in the bath. 
The ME-HEOM yield longer trapping times and indicate the importance of memory effects and correlations in order to maximize the efficiency in the FMO complex at physiological temperature.
Interestingly, the zero-shift energies of the FMO complex provide an almost optimal arrangement for funneling the energy flow to the reaction center at $T=300$~K.
Beyond the results for the FMO complex, our fast computational GPU-algorithm for the HEOM provides a robust and scalable way to treat bigger systems and allows us to calculate two-dimensional spectra of LHC, which requires to enlarge the dimension of the density matrix by taking into account double-excitonic states.

\section*{Acknowledgement}
This work has been supported by the DAAD project 50240755 and the Spanish MINCINN AI  DE2009-0088 (Acciones Integradas Hispano-Alemanas), the Emmy-Noether program of the DFG, KR~2889/2, the Spanish MICINN project FIS2010-18799 and the Ram{\'o}n y Cajal program.

\appendix

\section{Hybrid Markov-HEOM approach}\label{sec:markovif}

\subsection{Secular Redfield approximation}

We follow Ref.~\cite{Rebentrost2009b} and expand the phonon part $\mathcal{L}_{\rm phon}$ up to second
order in the exciton phonon coupling.  We use the Born-Markov and secular approximation to obtain
\begin{eqnarray}\label{eq:lab6}
-\frac{\rmi}{\hbar} \mathcal{L}_{\rm ex-phon}&=&\sum_{m,\omega}\rmi L_m(\omega)[V_{m}^\dag(\omega)V_{m}(\omega),\bullet]\nonumber\\
&&+\sum_{m,\omega}\rmi\gamma_m(\omega)\mathcal{D}(V_m(\omega)),
\end{eqnarray}
where $V_{m}=|m\rangle \langle m|$ stands for the exciton operators and $\mathcal{D}(V)\rho=V\rho V^\dag-\frac{1}{2}V^\dag V\rho-\frac{1}{2}\rho V^\dag V$. The
Lamb shift reads
\begin{equation}
 L_m(\omega)=\mbox{Im}\int_0^{\infty}{\rm d}t'\,e^{-i\omega t'}\langle u_m(t')u_m(0) \rangle_{\rm phon},
\end{equation}
and the decoherence rates are given by
\begin{equation}
 \gamma_m(\omega)=2\mbox{Re}\int_0^{\infty}{\rm d}t'\,e^{-i\omega t'}\langle u_m(t')u_m(0) \rangle_{\rm phon},
\end{equation}
with the phonon operators
$u_{m,{\rm phon}}= \big(\sum_i \hbar \omega_i d_i(b_i+b_i^\dag)\big)_m$.
The exciton operators
\begin{equation}
V_{m}(\omega)=\sum_{\omega,M,N}c_m^\ast(M)c_m(N)|M\rangle\langle N| \delta(\omega-E_M+E_N)
\end{equation}
are evaluated in the excitonic eigenbasis $|M\rangle=\sum_m c_m(M)|m\rangle$ with $\mathcal{H}_{\rm ex}|M\rangle=E_M|M\rangle$.
For simplicity we assume that the phonon environments of the individual
chromophores are uncorrelated. We additionally neglect the Lamb-type renormalization term.
For the explicit evaluation of the decoherence rates, we quantify the strength of the exciton-phonon coupling and
introduce a Drude-Lorentz spectral density
\begin{equation}
J_m(\omega)=2\lambda_m\frac{\omega \gamma_m}{\omega^2+\gamma_m^2}.
\end{equation}
The decoherence rates are then given by
\begin{equation}
\gamma(\omega)=
\left\{
\begin{array}{ll}
2\pi J(-\omega)(n(-\omega)+1), & \mbox{if }\omega<0 \\
2\pi\frac{k_BT}{\hbar}\frac{\mbox{\footnotesize d} J(\omega)}{\mbox{\footnotesize d}\omega}, &  \mbox{if }\omega=0 \\
2\pi J(\omega)n(\omega), & \mbox{if }\omega>0
\end{array}\right.
,
\end{equation}
where $n(\omega)=(\exp(\hbar\omega/k_B T)-1)^{-1}$
corresponds to the phonon statistics. Note that we neglect the index $m$ and use the same parameters for all sites.

We describe the radiative decay and trapping to the reaction center by a Lindblad ansatz
\begin{equation}\label{eq:lab2}
-\frac{\rmi}{\hbar}\mathcal{L}_{\rm ex-phot}=\sum_{m=1}^N\Gamma_{\rm phot}\mathcal{D}(|0\rangle\langle m|),
\end{equation}
and
\begin{equation}\label{eq:lab3}
-\frac{\rmi}{\hbar}\mathcal{L}_{\rm ex-RC}=\sum_{m=3}^4\Gamma_{\rm RC}\mathcal{D}(|RC\rangle\langle m|),
\end{equation}
with identical trapping rates for all sites $\Gamma_{\rm phot}=2\pi|\mu|^2$ and $\Gamma_{\rm RC}=2\pi|\mu_{RC}|^2$,  respectively.
\ref{eq:lab2} and \ref{eq:lab3} are derived by a master equation approach employing the rotating wave approximation for Hamiltonians $\mathcal{H}_{\rm phot}^0$ and $\mathcal{H}_{\rm phot}^{\rm RC}$ respectively. The
strength of the exciton-photon coupling is defined by a constant spectral density $J(\omega)=2\pi$. We further assume that
the photon field cannot create excitations. The same holds for the trapping to the reaction center, as no backward energy flow to the system is allowed. This is equivalent to previous works, where trapping and exciton recombination are included in the Hamiltonian in the form of anti-Hermitian parts \cite{Rebentrost2009a}.\\
The equation of motion for the density operator in secular Redfield form reads
\begin{equation}
\frac{\rmd}{\rmd t}\rho(t)=-\frac{\rmi}{\hbar} \left( 
 \mathcal{L}_{0}
+\mathcal{L}_{\rm ex-phon}
+\mathcal{L}_{\rm ex-phot}
+\mathcal{L}_{\rm ex-RC}
\right)\rho(t).
\end{equation}

\subsection{Combined master-equation-HEOM approach (ME-HEOM)}

For the FMO complex, the coupling of the excitons to the phonon environment is large and the reorganization process occurs on
a time scale which is comparable to the system dynamics. The secular Redfield approximation is not valid for the coupling to the phonon bath and a non-perturbative treatment is required.
We will follow the derivation in Ref.~\cite{Ishizaki2009c} and introduce a set of hierarchically coupled equations.
For trapping-time and efficiency calculations we introduce slight modifications and in particular
we include the coupling to the reaction center and the radiative decay. We derive a combined
ME-HEOM approach which treats the exciton-phonon coupling exactly, whereas the
leakage to the reaction center and exciton ground state is described in the Born-Markov limit.

We start with the Liouville equation for the total density operator \ref{eq:L} and assume that the total density operator factorizes \ref{eq:R}. In the interaction picture with
\begin{equation}
\mathcal{H}_0=
 \mathcal{H}_{\rm ex}
+\mathcal{H}_{\rm trap}
+\mathcal{H}_{\rm phon}
+\mathcal{H}_{\rm phot}^{0}
+\mathcal{H}_{\rm phot}^{\rm RC},
\end{equation}
where we denote operators with
\begin{equation}
\tilde{\mathcal{O}}(t)=
\rme^{\rmi \mathcal{H}_0 t /\hbar}
\mathcal{O}
\rme^{-\rmi \mathcal{H}_0 /\hbar},
\end{equation}
the Liouville equation reads
\begin{eqnarray}
\dt\tilde{R}(t)&=&-\frac{\rmi}{\hbar}[
 \tilde{\mathcal{H}}_{\rm ex-phon}
+\tilde{\mathcal{H}}_{\rm ex-phot}
+\tilde{\mathcal{H}}_{\rm ex-RC}
,\ \tilde{R}(t)]\nonumber\\
=&-&\frac{\rmi}{\hbar}\left(
 \tilde{\mathcal{L}}_{\rm ex-phon}
+\tilde{\mathcal{L}}_{\rm ex-phot}
+\tilde{\mathcal{L}}_{\rm  RC}
\right)\tilde{R}(t).
\end{eqnarray}
After formal integration and tracing out the bath degrees of freedom $\alpha=\{\rm phon, phot^0, phot^{RC}\}$ we get a formal solution for the reduced density operator describing
the exciton degrees of freedom
\begin{equation}
\tilde{\rho}(t)=\tilde{\mathcal{U}}(t)\tilde{\rho}(0)
\end{equation}
with time evolution operator
\begin{eqnarray}
\tilde{\mathcal{U}}(t)=\mbox{Tr}_\alpha\big( T_+
&{\rm e}^{-\frac{\rmi}{\hbar}\int_0^t{\rm d}s\,(
 \tilde{\mathcal{L}}_{\rm ex-phon}(s)+\tilde{\mathcal{L}}_{\rm ex-phot}(s)+ \tilde{\mathcal{L}}_{\rm ex-RC}(s))
}\nonumber \\
&\times \rho_{\rm phon}\otimes\rho_{\rm phot}^0\otimes\rho_{\rm phot}^{\rm RC}\big).
\end{eqnarray}
We make use of the Gaussian nature of the harmonic baths to reduce the bath expectation values to two-time
correlation functions. Hence the influence of the environment is characterized by the
symmetrized correlation
\begin{equation}
S_{m,\alpha}(t)=\frac{1}{2}\langle[\tilde{u}_{m,\alpha}(t),\tilde{u}_{m,\alpha}(0)]_+ \rangle,
\end{equation}
and the response function
\begin{equation}
\chi_{m,\alpha}(t)=\frac{1}{2}\langle[\tilde{u}_{m,\alpha}(t),\tilde{u}_{m,\alpha}(0)] \rangle,
\end{equation}
where $u_{m, \rm phot}=\sum_{\nu} \mu_{m}^\nu (a_\nu+a^\dag_\nu)$.
We assume that each site is coupled to an independent phonon bath and that there are no correlations between the radiative decay and trapping at different sites. 
For the exciton-phonon coupling we employ a Drude-Lorentz spectral density
\begin{equation}\label{eq:gamma}
J_m(\omega)=2\lambda_m\frac{\omega \gamma_m}{\omega^2+\gamma_m^2},
\end{equation}
and obtain, in the high temperature limit
\begin{eqnarray}
S_m(t)\simeq\frac{2 \lambda_m}{k_bT}e^{-\gamma_m t}, \\
\chi_m(t)= 2 \lambda_m \gamma_me^{-\gamma_m t}.
\end{eqnarray}
The parameter $\gamma_{m}$ describes the time scale of correlations in the vibrational environment of the protein. Note that as we consider identical couplings for all sites, the notation is simplified in the main text and the subindex $m$ is removed from the time correlation scale of the bath $\gamma$. \\
The time evolution operator becomes
\begin{equation}
 \tilde{\mathcal{U}}(t)=T_+\prod_{m=1}^N
\rme^{\int_0^{t}{\rm d}s\,\tilde{W}_{m,\rm phon}(s)}
\prod_{m=1}^N
\rme^{\int_0^{t}{\rm d}s\,\tilde{W}_{m,\rm phot^0}(s)}
\prod_{m=3}^4
\rme^{\int_0^{t}{\rm d}s\,\tilde{W}_{m,\rm phot^{RC}}(s)}
\end{equation}
with
\begin{equation}\label{eq:4a}
 \tilde{W}_{m,\alpha}=-\frac{1}{\hbar^2}\int_0^t{\rm d}s\,
 \tilde{V}_{m,\alpha}(t)^\times[S_{m,\alpha}(t-s)\tilde{V}_{m,\alpha}(s)^{\times}-\rmi\frac{\hbar}{2}\chi_{m,\alpha}\tilde{V}_{m,\alpha}(s)^\circ].
\end{equation}
We denote the commutation relations by $\mathcal{O}^\times f=[\mathcal{O},f]$ and $\mathcal{O}^\circ f=[\mathcal{O},f]_+$.
The time evolution of the reduced density matrix is given by
\begin{equation}\label{eq:4}
\dt \tilde{\rho}(t)=T_+\Big(\sum_{m=1}^7 \tilde{W}_{m,\rm phon}(t)+\sum_{m=1}^7 \tilde{W}_{m,\rm phot^0}(t)
+\sum_{m=3}^4 \tilde{W}_{m,\rm phot^{RC}}(t)\Big)\tilde{\rho}(t).
\end{equation}
Note that due to the time ordering operator affecting the integration in Eqs.~(\ref{eq:4a},\ref{eq:4}) is time non-local. In the following we treat the exciton-photon and
exciton-reaction center couplings in the Born-Markov limit. That is, the time non-local operators $T_+\sum_{m=1}^7\tilde{W}_{m,\rm phot^0}(t)$ and
$T_+\sum_{m=3}^4\tilde{W}_{m,\rm phot^{RC}}(t)$ are replaced by their time-local Born-Markov limit $\mathcal{L}_{\rm ex-phot}$ and $\mathcal{L}_{\rm ex-RC}$
defined in Eqs.~(\ref{eq:lab2},\ref{eq:lab3}), respectively. Eq.~(\ref{eq:4}) finally reduces to
\begin{equation}\label{eq:5}
 \dt \tilde{\rho}(t)=
 -\frac{\rmi}{\hbar}\mathcal{L}_{\rm phot}\tilde{\rho}(t)
 -\frac{\rmi}{\hbar}\mathcal{L}_{\rm RC}\tilde{\rho}(t)
+T_+\sum_{m=1}^7 \tilde{W}_{m,\rm phon}(t)\tilde{\rho}(t)
.\end{equation}
We define auxiliary operators
\begin{equation}
 \tilde{\sigma}^{(n_1, ...,n_7)}(t)=T_+\prod_{m,k,l}\Big(\int_0^t {\rm d}s\,
{\rm e}^{-\gamma_m(t-s)\tilde{\theta}_m(s)}\Big)^{n_m}
{\rm e}^{\int_0^t{\rm d}s\,\tilde{W}_{m,\rm phon}(s)}
{\rm e}^{\int_0^t{\rm d}s\,\tilde{W}_{k,\rm phot^0}(s)}
{\rm e}^{\int_0^t{\rm d}s\,\tilde{W}_{l,\rm phot^{RC}}(s)} \label{eq:M}
\end{equation}
with  
\begin{eqnarray}
\tilde{\theta}_m(s)&
=&\rmi\Big(\frac{2\lambda_m}{k_B T\hbar^2}\tilde{V}_{m,\rm phon}^{\times}(s)
-\rmi\frac{\lambda_m}{\hbar}\gamma_m \tilde{V}_{m,\rm phon}^{\circ}(s)\Big),\nonumber \\
\tilde{\sigma}^{(0,..,0)}(t)&=&\tilde{\rho}(t),
\end{eqnarray}
and rewrite the time non-local effects into hierarchically coupled equations of motion
\begin{eqnarray}\label{eq:labsevena}
 \dt\rho(t)=
&-&\frac{\rmi}{\hbar}
\Big(\mathcal{L}_{\rm ex}+\mathcal{L}_{\rm phot}+\mathcal{L}_{\rm RC}\Big)\rho(t)\nonumber\\
&+&\sum_m
\rmi V_{m,\rm phon}^{\times}
\sigma^{(n_1, ...,n_m+1,...,n_7)}(t)
\end{eqnarray}
with
\begin{eqnarray}\label{eq:lab8}
& \dt&\sigma^{(n_1, ...,n_7)}(t)=\\
&=&\big[
-\frac{\rmi}{\hbar}(\mathcal{L}_{\rm ex}
+\mathcal{L}_{\rm phot}
+\mathcal{L}_{\rm RC})
+\sum_m n_m\gamma_m\big]\sigma^{(n_1, ...,n_7)}(t)\nonumber\\
&&+\sum_m
\rmi V_{m,\rm phon}^{\times}
\sigma^{(n_1, ...,n_m+1,...,n_7)}(t)\nonumber \\
&&+\sum_mn_m\theta_m\sigma^{(n_1, ...,n_m-1,...,n_7)}(t)\nonumber ,
\end{eqnarray}
where we again have used the Born-Markov limit for the trapping and radiative decay.
The auxiliary operators keep track of the memory effects of the bath and account for the removal of the
reorganization energy. The $\sigma$-matrices are initially set to zero. For a sufficiently large $N_{\rm max}=\sum_m n_m$, the diagonal coupling in Eq.~(\ref{eq:lab8}) becomes the dominant term and we can truncate the hierarchy.

\section{Algorithm for implementing the hierarchical method on graphics processing units}\label{sec:gpu}

For the large reorganization energies typically found in LHC one needs to go beyond  the Born-Markov approach and to consider non-local temporal effects. We do this by solving the system dynamics within the hierarchical approach shown in the previous section. The method requires considerable
memory and computational efforts and a large number of auxiliary
matrices is needed to store the time non-local bath properties.
Since all auxiliary matrices have to be accessed to perform the next
propagation step, the huge communication overhead renders conventional
parallelization schemes, where distributed computing nodes are connected 
by Ethernet \cite{Struempfer2009a}, ineffective.
GPUs have the twofold advantage of a fast memory
bandwidth and the availability of several hundred stream processors. By
assigning one stream-processor to each auxiliary matrix we obtain a
speedup of the hierarchical method by the number of processors. The
numerical calculations in this manuscript are performed on a NVIDIA
Fermi C2050 GPU with $448$ processors (1.15 GHz) and 3 gigabytes of 
ECC-protected on-board memory.
\begin{table}[t]
\begin{center}
 \begin{tabular}{c c c c c c}
\hline
     $N_{\rm max}$  & \#$\sigma$-matr. & CPU & GPU & speed up & GPU utilization \\
\hline
4 &  330 & 120 s& 1 s & $\times$120 & 22\% \\
6 & 1\,716 & 676 s& 3 s & $\times$225 & 56\% \\
8 & 6\,435 & 2\,636 s& 7 s & $\times$376 & 82\% \\
10 & 19\,448 & 8\,275 s& 19 s & $\times$435 & 93\% \\
12 & 50\,388 & 21\,972 s& 48 s & $\times$458 & 97\%
\end{tabular}
\caption{\label{tab:GPUvsCPUTimeingtab1}
Comparison of CPU and GPU computation time of the population dynamics of the isolated FMO complex.
We propagate 1000 time steps, the GPU (NVIDIA C2050) calculation are performed in single precision.
Double precision (not required here for converged results) increases the GPU computation time by a factor of two.
}
\end{center}
\end{table}
The first step of the algorithm initializes the system of the $\sigma$-matrices of the hierarchy. With increasing truncation $N_{\rm max}$, the total number of $\sigma$-matrices grows factorially  $N_{\rm tot}=(N+N_{\rm max})!/(N!N_{\rm max}!)$, where $N$ corresponds to the number of sites \cite{Ishizaki2009a}.  As shown in \ref{tab:GPUvsCPUTimeingtab1}, the calculation of a population dynamics of the FMO complex with $N=7$, $N_{\rm max}=12$ requires already 50\,388 matrices whereas 330 matrices are sufficient for a truncation at $N_{\rm max}=4$.  The memory of the $\sigma$-matrices is allocated on the graphics-board and initialized to zero. It is not necessary to transfer the $\sigma$-matrices to the main-processor memory at any time during the calculation. The only memory transfer between CPU and GPU involves the $N\times N$ entries of the reduced density operator $\rho$. To advance the propagation one time-step in eq.~(\ref{eq:lab8}) requires a large number of matrix multiplications. Each single $\sigma$-matrix is connected to $2N$ neighbors, these connections are stored in GPU memory in a linked-list. The GPU uses 448 cores in parallel with fast GPU memory transfer and thus provides an immense reduction of the computation time up to a factor of $458$ for the matrix multiplications. For benchmarking the algorithm, we propagate 1000 time steps using a 4th order Runge-Kutta integrator. For the final output into files requires a short memory transfer from the GPU to the CPU.

In \ref{tab:GPUvsCPUTimeingtab1} we summarize the computational speed-up of the C2050-GPU compared to a standard CPU (Intel 2.40GHz).
The GPU computation is performed using single precision, which yields sufficient accuracy for the problem at hand. For the population dynamics of the FMO complex using $\lambda=35$~cm$^{-1}$, $\gamma^{-1}=166$~fs, temperature of 300 K, propagation time of 10~ps with step size $\Delta t=10$~fs and truncation $N_{\rm max}=12$ the populations are accurate within single precision to six digits 
$|\rho_{ii}^{\rm single}(t)-\rho_{ii}^{\rm double}(t)|<5\times10^{-7}$.
This switch from single for double precision increases the computation time approximately by a factor of two on the C2050-GPU.

\end{document}